\documentclass[amsmath,amssymb,floatfix,lengthcheck,preprintnumbers,prl,showpacs,superscriptaddress,nofootinbib,twocolumn,aps]{revtex4-1}
\usepackage{graphicx}
\usepackage{xcolor,ulem}
\usepackage{mathtools}

\newcommand{\nn}{\nonumber}
\newcommand{\eq}[1]{eq.~\ref{#1}}

\newcommand{\fig}[1]{figure~\ref{#1}}

\newcommand{\refcite}[1]{ref.~\cite{#1}}

\long\def\/*#1*/{}
\definecolor{red}{rgb}{1.0, 0, 0}

\newcommand{\op}[0]{\ensuremath{\mathcal{O}}}
\newcommand{\Rop}[0]{\ensuremath{\mathcal{R}}}
\newcommand{\symop}[0]{\ensuremath{\mathcal{A}}}

\begin{document}

\title{Nonperturbative Renormalization of Operators in Near-Conformal Systems Using Gradient Flows}

\author{Andrea Carosso}\email{Andrea.Carosso@colorado.edu}
\affiliation{Department of Physics, University of Colorado, Boulder, Colorado 80309, USA}

\author{Anna Hasenfratz}\email{Anna.Hasenfratz@colorado.edu}
\affiliation{Department of Physics, University of Colorado, Boulder, Colorado 80309, USA}

\author{Ethan T.~Neil}\email{ethan.neil@colorado.edu}
\affiliation{Department of Physics, University of Colorado, Boulder, Colorado 80309, USA}
\affiliation{RIKEN-BNL Research Center, Brookhaven National Laboratory, \\ Upton, New York 11973, USA}

\date{June 4, 2018}

\begin{abstract}

We propose a continuous real space renormalization group transformation based on gradient flow, allowing for a numerical study of renormalization without the need for costly ensemble matching.  We apply our technique in a pilot study of SU$(3)$ gauge theory with $N_f = 12$ fermions in the fundamental representation, finding the mass anomalous dimension to be $\gamma_m = 0.23(6)$, consistent with other perturbative and lattice estimates.  We also present the first lattice calculation of the nucleon anomalous dimension in this theory, finding $\gamma_N = 0.05(5)$.

\end{abstract}

\preprint{}

\maketitle

\paragraph{Introduction --}  Conformal field theories describe a number of important physical systems.  In lower dimensions, they  describe the critical behavior of a wide array of models.  The use of conformal field theories in four dimensions ranges from composite Higgs models to $\mathcal{N}=4$ super-Yang-Mills theory, and through the AdS/CFT correspondence  to the definition of quantum gravity and beyond. 

Non-perturbative techniques to obtain the spectrum of operator dimensions are essential to study the full range of strongly-coupled conformal field theories.  These include the conformal bootstrap \cite{ElShowk:2012ht,Poland:2016chs,Poland:2018epd}, radial quantization \cite{Brower:2012vg,Brower:2016vsl,Brower:2018szu}, and Monte Carlo Renormalization Group (MCRG) \cite{Swendsen:1979gn,Bowler:1984hv,Hasenfratz:2011xn}.  MCRG in particular is exact and non-perturbative, but conventional approaches are limited by the requirement of matching ensembles over large, discrete changes in scale.

Gradient flow (GF)~\cite{Narayanan:2006rf,Luscher:2009eq} is a continuous, invertible field transformation that systematically suppresses high momentum modes, allowing for the definition of renormalized quantities non-perturbatively. Formally, it is similar to the coarse-graining step of momentum space  RG transformations, such as
Wilson and Kogut's ``incomplete" integration as described in \refcite{Wilson:1973jj}, Polchinski's ``smooth cutoff''~\cite{Polchinski:1983gv}, and
the coarse-graining function of functional RG~\cite{Berges:2000ew,Bagnuls:2000ae}.
Analogously, in position space the flowed fields could be considered as RG blocked fields.  Significant recent work has gone into exploring the connections between GF and RG \cite{Aoki:2014dxa,Aoki:2015snd,Monahan:2015lha,Fujikawa:2016qis,Capponi:2016yjz,Makino:2018rys,Abe:2018zdc}.

Despite these similarities, GF itself is not an RG transformation, as it lacks two essential steps which are necessary for the existence of a fixed point (FP):
\begin{enumerate}
\item GF does not include a rescaling that would restore the momentum cutoff to its original value.
\item Linear GF, like the fermion flow~\cite{Luscher:2013cpa} or the free scalar flow~\cite{Monahan:2015lha,Fujikawa:2016qis,Capponi:2016yjz} does not uniquely fix the normalization of the field. Linear RG transformations do not have a FP unless the blocked fields are normalized by $b^{ \eta/2}$, where $b$ is the scale change and $\eta/2$ is the anomalous dimension of the field.
\end{enumerate}

RG flow may also be thought of as a map on the ``theory space'' of couplings $g(\mu)$ such that the combination of a dilatation (scale transformation) by scale factor $b$ and a rescaling of the couplings $g(\mu) \rightarrow g(b\mu)$ leaves all physical predictions of the theory unchanged \cite{Wilson:1973jj,Gukov:2015qea}.  The required dilatation is the first step above.  The second step is simply the requirement that the normalization of the kinetic terms for all fields should be unchanged.

In numerical simulations, the renormalization of the fields can be included by explicit calculation of the wavefunction renormalization using an exactly conserved current. We show that by considering correlation functions at long distances the dilatation transformation can be incorporated without  an explicit rescaling step.

The use of GF provides several advantages over conventional MCRG techniques. Since GF is a continuous transformation, the corresponding RG can also be continuous, greatly improving the predictive power of the method and avoiding the costly matching over discrete scale changes.  In addition, GF gives a straightforward definition of blocked lattice fermion fields that allows the evaluation of the blocked fermion correlation functions without requiring the knowledge of the blocked action. This opens a greater range of theories to be studied.

In the literature on continuum RG, ``gradient flow'' often has a specific connotation in terms of the strongest form of the $c$-theorem conjecture \cite{Barnes:2004jj}, namely that RG flow is precisely the gradient flow of the $c$-function.  We do not claim any result on the $c$-theorem here.  For our purposes, we only require a flow which suppresses high-momentum modes of quantum fields. Changing the flow definition changes the  renormalization scheme.

In this paper, we show how continuous flow can be used to define an RG transformation, focusing on the use of flowed correlation functions at large separation to extract anomalous dimensions.  We then apply our method to determine the anomalous dimension of various fermionic operators in SU(3) gauge theory with 12 fundamental flavors. While the infrared properties of this model are still controversial~\cite{Fodor:2011tu,Cheng:2013xha,Cheng:2013xha,Lombardo:2014pda,Lin:2015zpa,Fodor:2016zil,Hasenfratz:2016dou,Fodor:2017gtj} our analysis is consistent with the existence of an infrared fixed point in the lattice theory with staggered fermions - although we expect the proposed method to predict the scale-dependent renormalization factors even in the absence of such a fixed point.  We find the mass anomalous dimension as $\gamma_m=0.23(6)$ and the anomalous dimension of the nucleon as $\gamma_N = 0.05(5)$ - the first non-perturbative prediction of this quantity.  These values are consistent both with other perturbative~\cite{Ryttov:2010iz,Pica:2016rmv,Ryttov:2017kmx,Gracey:2018oym} and numerical~\cite{Cheng:2013bca,Cheng:2013xha,Lombardo:2014pda,Giedt:2015alr} predictions.

\section {Gradient flow and RG transformations }

In this section we denote all fields (scalar, fermion, or gauge)  by $\phi$ and the corresponding gradient-flowed fields by $\phi_t$, where $t$ is the dimensionless flow parameter.  The specific form of the flow is not important, we require only that it suppresses high momentum modes above $\Lambda_{0}/\sqrt{t}$, where $\Lambda_0$ is the cutoff scale; for a lattice cutoff $\Lambda_0 \sim 1/a$, where $a$ is the lattice spacing.  The identification of $\sqrt{t}$ with the smearing range corresponds to the conventional definition of gradient flow \cite{Luscher:2010iy}.

 To begin with, we assume the existence of an IR-conformal (critical) FP in the theory to be studied.  Working in the basin of attraction, the fixed point can be characterized by a set of scaling fields with irrelevant couplings we denote generally as $g$.  We also include a relevant coupling $m$ that breaks the conformal symmetry, though at the end we will set $m=0$ and study the system at criticality.  The couplings $g$ and $m$ are defined to vanish at the FP.

We consider an arbitrary local operator projected to zero spatial momentum on timeslice $x_0$, $\op(x_0) = \int d\boldsymbol{x}\ \op(\phi; x)$.  Under an RG transformation which changes the lattice cutoff as $a \to a' = ba$ , $b>1$, the couplings transform with their corresponding scaling dimensions, and the two-point correlation function of $\op$ at distance $x_0 \gg a' =  ba $ transforms as \cite{cardy1996scaling,DelDebbio:2010ze}
\begin{equation}
\langle \op(0) \op(x_0) \rangle_{g,m} = b^{-2\Delta_\op} \langle \op(0) \op(x_b) \rangle_{g',m'},
\label{eq:CO_standard}
\end{equation}
where all physical quantities (except the lattice cutoff $a'$) on the right-hand side have been rescaled by $b$, so that $x_b \equiv x_0 / b$.  Here $\Delta_\op = d_\op + \gamma_\op$ is the scaling dimension of the operator $\op$, which is divided into its canonical dimension $d_\op$ and anomalous dimension $\gamma_\op$.

The RG-transformed correlator on the right-hand side of eq.~\ref{eq:CO_standard} is evaluated with respect to the action  with the transformed couplings $g'$ and $m'$.  Rather than attempting to study the RG flow of the action itself directly, we may use the principle of MCRG~\cite{Swendsen:1979gn}, which states that RG transformation and the generation of a Monte Carlo ensemble from  the action are commuting operations.  In other words, we can RG transform the ensemble of fields generated using the original action, rather than attempting to construct the new action explicitly, i.e.
  \begin{equation}
 \langle \op(0) \op(x_b)\rangle_{g^\prime,m^\prime} =  \langle \op_b(0) \op_b(x_b)\rangle_{g,m},
 \label{eq:MCRG} 
 \end{equation}
where $\op_b$ is the same local operator constructed from the blocked fields $\Phi_b$ rather than the original $\phi$.  We emphasize that the RG arguments leading to Eq.~\ref{eq:CO_standard} are valid only when the blocked operators are well separated and the blocked fields making up $\op_b(0)$ and $\op_b(x_b)$ do not overlap.  

We define the blocked fields $\Phi_b$ in terms of flowed fields as
\begin{equation}
\Phi_b (x_b)\equiv b^{\Delta_\phi} \phi_t( bx_b ),
\label{eq:block_flow}
\end{equation}
where the exponent $\Delta_\phi = d_\phi + \eta/2$ is fixed by requiring the $\phi$ two-point correlator at criticality, which (in infinite volume) takes the form of a power law in $x_0$, to be unchanged by RG transformation as described by Eq.~\ref{eq:CO_standard}, i.e. $\eta/2$ is the anomalous dimension of the field $\phi$.  At this point the flow time is independent of the RG scale change but it is natural to keep the two related asymptotically as $\sqrt{t} \propto b$, since a substantial mismatch between blocking radius and scale change could distort the RG transformation and obstruct the existence of an RG fixed point; this is known to be the case for decimation, which corresponds to $\sqrt{t} \ll b$.

Combining eqs.~\ref{eq:CO_standard} and \ref{eq:MCRG} with eq.~\ref{eq:block_flow} and the identification of $b$ yields the result 
\begin{equation}
\frac{ \langle \op_t(0) \op_t(x_0) \rangle}{ \langle \op(0) \op(x_0) \rangle} = b^{2\Delta_\op - 2n_\op \Delta_\phi} \propto t^{n_{d,\op} + \gamma_\op - n_\op \eta / 2} \label{eq:ratio2}
\end{equation}
where $\op_t = \op(\phi_t)$ is the operator in terms of the GF fields, $n_\op$ is the number of $\phi$ fields in operator $\op$, and $n_{d,\op} = d_{\op} - n_\op d_\phi$ is the number of derivatives appearing in $\op$.  We emphasize that the flowed and unflowed operators are evaluated at the same lattice distance $x_0/a$; the transformation of the lattice cutoff $a \rightarrow ba$ is exactly compensated by the rescaling step of the RG transformation.

Eq.~\ref{eq:ratio2} allows the determination of the anomalous dimension of $\op$ directly from the flow-time dependence of its correlation functions.  However, it is sometimes numerically advantageous to compute expectation values with only a single flowed operator; for example, if $\phi$ is a fermion then the costly ``adjoint flow'' \cite{Luscher:2013cpa} is required to compute $\langle \op_t(0) \op_t(x_0)\rangle$. The flowed operator $\op_t(x_0)$ is a combination of local fields in the vicinity of $x_0$ with coefficients that depend on the flow time,
\begin{equation}
\op_t(x_0) \sim \sum_{k} c_k(t) \op(x_0 + \delta x_k),
\end{equation}
where the coefficients $c_k(t)$ are exponentially suppressed for distances beyond the  smearing radius of the flow $\propto \sqrt{t}$.  The condition $x_0 \gg ba \propto a \sqrt{t}$ required for \eq{eq:CO_standard} thus implies that $\delta x_k \ll x_0$. Expanding in $a \sqrt{t} / x_0$ and using translation invariance, it is straightforward to show that
\begin{equation}
\frac{ \langle \op(0) \op_t(x_0) \rangle}{ \langle \op(0) \op(x_0) \rangle} \propto t^{n_{d,\op} / 2 + \gamma_\op/2 - n_\op \eta/4} + O(a\sqrt{t}/x_0) \label{eq:ratio}
\end{equation}
where the dependence on $t$ is the square root of the dependence in \eq{eq:ratio2}.

We can use \eq{eq:ratio} to determine the field anomalous dimension $\eta$, as long as we can identify some local operator $\symop$ which is protected from renormalization by a symmetry of the theory, and therefore has $\gamma_{\symop} = 0$.  Once $\eta$ is determined, any other anomalous dimension can be predicted.  Alternatively, we may construct a double ratio of the form
\begin{eqnarray}
\Rop_{\op}(t,x_0) &=& \frac{ \langle \op(0) \op_t(x_0) \rangle}{ \langle \op(0) \op(x_0) \rangle} \label{eq:ratio_full}
\Big( \frac{ \langle \symop(0) \symop(x_0) \rangle}{ \langle \symop(0) \symop_t(x_0) \rangle} \Big)^{n_\op/n_\symop}  \\ 
&=& b^{\Delta_\op - (n_\op / n_\symop) d_\symop}  \nn \\
&\propto& t^{\gamma_\op/2 + \delta/2}, \quad x_0 \gg a\sqrt{t}  \nn
\end{eqnarray}
which cancels the anomalous dimension $\eta$ directly, leaving only the desired anomalous dimension $\gamma_\op$ and some possible residual dependence on the canonical dimensions of $\op$ and $\symop$ through $\delta \equiv d_\op - (n_\op / n_\symop) d_\symop$.  If the operators contain no derivatives then $\delta = 0$; this will be the case for all operators we consider in our numerical study. 

Eq.~\ref{eq:ratio_full} is valid only on the critical $m=0$ surface and at sufficiently large flow times such that the linear basin of attraction of the IR-stable fixed point has been reached.  Otherwise, we expect the predicted $\gamma_O$ from eq.~\ref{eq:ratio_full} to show additional dependence on $t$ coming from irrelevant operators.  In practice, the flow time $t$ which can be reached is limited by the finite lattice volume.

\paragraph{Finite volume corrections -- } 


Consider the  ratio $\Rop(g^\prime,t,L)$ on lattice size $L$ at flow time $t$ starting at bare coupling $g^\prime$, and  compare it to the ratio on lattice size $sL$ at flow time $s^2 t$ starting at bare coupling $g$.  In the basin of attraction of the FP, the coupling $g^\prime$ can be adjusted so that the two flows end at the same physical point, so that eq.~\ref{eq:ratio_full} now predicts
\begin{equation}
\Rop_{\op}(g^\prime,t,L) = s ^{-\gamma_\op} \Rop_{\op}(g,s^2t,sL)\, .
\end{equation}
Applying this relation twice and expanding the right hand side around $g$
\begin{eqnarray}
\Rop_{\op}&&(g,s^2 t,s^2 L) = \Rop_{\op}(g,s^2 t,s L) +  \nn \\ 
&& s ^{\gamma_\op} \big( 
\Rop_{\op}(g,t,sL) - \Rop_{\op}(g,t,L) \big) + O(g^\prime - g) \, .
\label{eq:finite_vol_corr}
\end{eqnarray}
Eq.~\ref{eq:finite_vol_corr} predicts the ratio $\Rop(g)$ on volume $s^2 L$ in terms of ratios on smaller volumes, plus a correction term $O(g^\prime - g)$.  We will absorb the latter term as a $g$ dependent correction and assume that the ratio on $s^2 L$ volumes approximates infinite volume.  Assuming that conformal symmetry is broken only by the finite number of spatial lattice points $L$, we expect finite volume corrections to depend only on the dimensionless ratio $b/L$, and thus on the flow time as $\sqrt{t}/L$.

\section{Details of lattice simulations } 

To test our proposed method numerically, we carry out a pilot study of SU(3) gauge theory with $N_f = 12$ degenerate fermions in the fundamental representation.  We use a set of gauge configurations that were originally generated for finite-size study of this system~ \cite{Cheng:2013xha} using a plaquette gauge action and  nHYP-smeared staggered fermions~\cite{Hasenfratz:2001hp,Hasenfratz:2007rf}.  Further details on the lattice action  can be found in Refs.~\cite{Cheng:2011ic,Cheng:2013eu,Cheng:2013bca,Cheng:2013xha}.   We consider five values of the bare gauge coupling $\beta=4.0,5.0,5.5,5.75$ and $6.0$, analyzing 46 and 31 configurations on lattice volumes of $24^3\times 48$ and $32^3\times 64$, respectively.  The fermion mass is set to $m = 0.0025$, small enough that we expect the breaking of scale invariance to be dominated by the finite spatial extent $L$.

We consider only fermionic operators, and use  the  axial charge $A^4$  for our conserved operator $\mathcal{A}$. Since staggered fermions have a remnant U(1) symmetry, it is straightforward to construct a conserved axial charge operator with $Z_A=1$ \cite{Aoki:1999av}.    We use on-site staggered operators for the pseudoscalar, vector, and nucleon, and a 1-link operator for the axial charge states.   Our individual correlators are consistent with simple exponential decay, although we cannot rule out a functional dependence that includes a Yukawa-like power law correction \cite{Ishikawa:2013tua}.


 Following \refcite{Luscher:2013cpa}, we adopt non-linear Wilson flow for the gauge fields and linear fermion flow.  We consider 10 flow time values between $1.0 \le t/a^2 \le 7.0$ (note that the flow range is $\sqrt{8t}$.) The strong correlations in GF lead to very small statistical errors in the flow-time dependence.  


\paragraph{Analysis -- }

\begin{figure}
\includegraphics[width=0.48\textwidth]{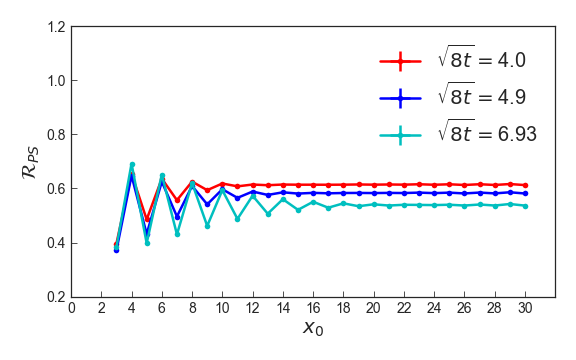}
\caption{Dependence of the correlator ratio $\Rop_P$ on source-sink separation $x_0$ and flow scale $\sqrt{8t}$.  For each value of $\sqrt{8t}$, a stable plateau in $R_P$ is seen for $x_0 \gtrsim 2\sqrt{8t}$.  The results shown here are on $32^3\times 64$ volumes at $\beta=5.75$. \label{fig:corr-flow}}
\end{figure}
In the following, we work in lattice units.  The ratio given in \eq{eq:ratio_full} should be independent of $x_0$ at large $x_0$, as long as the operator $\op$ has well defined quantum numbers. At distances comparable to the flow range, $x_0 \lesssim \sqrt{8t}$, the flowed operators overlap and the ratios could have non-trivial and non-universal structure. Since we are using staggered fermions where the action has oscillating phase factors, in the small $x_0$ region we observe significant oscillation, as shown in  \fig{fig:corr-flow} for the $\gamma_5$ pseudoscalar operator that does not have a partner in the channel. The width of the oscillation is about $2\sqrt{8 t}$, after which a stable plateau develops.  The decrease in the value of the plateau as the flow time increases predicts the anomalous dimension of the pseudoscalar operator.

We work directly with the ratio $\Rop(t)$ of eq.~\ref{eq:ratio_full}, and do not attempt to extrapolate the fermion field exponent $\eta$ (obtained from using $\symop$ in eq.~\ref{eq:ratio})  to the infrared limit, as it shows much stronger finite-volume and bare coupling dependence than the full operator ratios.  At fixed $t$ and $\beta$ we typically find $\eta \lesssim 0.1$.  

As a consistency check we consider the vector operator, but find large systematic effects due to oscillation; although we cannot quote a precise extrapolated value, we generally find the associated anomalous dimension consistent with zero as expected.

We predict the anomalous dimension as a function of $t$ by comparing the ratios at consecutive $(t_1,t_2)$ flow time values
\begin{equation}
\gamma_\op (\beta, \bar{t}, L) = 
  \frac{ \rm{log}(\Rop_\op(t_1,\beta,L) / \Rop_\op(t_2,\beta,L)) } { \rm{log}(\sqrt{t_1}/\sqrt{t_2}) }
\end{equation}
where $\bar{t}=(t_1+t_2)/2$. The mass anomalous dimension is predicted by considering the pseudoscalar operator, recalling that   $\gamma_m = -\gamma_{S}  = - \gamma_{PS} $. We estimate the  finite volume corrections  by \eq{eq:finite_vol_corr}, estimating $\gamma_m$ iteratively. We have numerical data on $24^3\times 48$ and $32^3\times 64$ volumes  so $s=32/24$,  and \eq{eq:finite_vol_corr} increases the effective volume to $42.66$.

\begin{figure}
\includegraphics[width=0.48\textwidth]{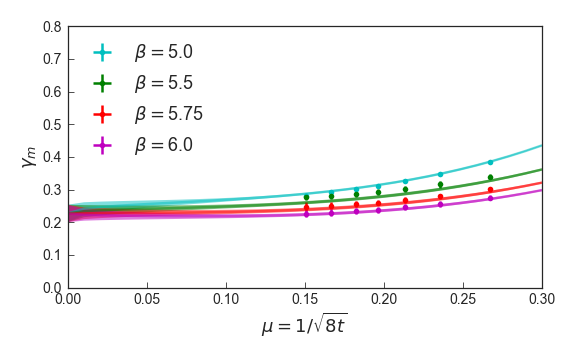}
\caption{Extrapolation of the mass anomalous dimension $\gamma_m$ to the infrared limit, as described in the text. \label{fig:gamma-M}}
\end{figure}

In \fig{fig:gamma-M} we show the infinite volume estimated $\gamma_m$ as a function of $\mu \equiv 1/\sqrt{ 8 \bar{t}}$. 
There is significant dependence on the bare gauge coupling $\beta$ and also on the flow time $t$, as expected in a slowly running system.  We extrapolate to the $t \to \infty$ limit as
\begin{equation}
\gamma_m(\beta,t) = \gamma_0 + c_\beta t^{\alpha_1} + d_\beta t^{\alpha_2}
\label{eq:extrapolation}
\end{equation}
motivated by the expectation that the correction terms should be due to the slowly evolving irrelevant couplings,  associated with higher-dimensional operators that can mix with the operator of interest.. Based on Refs.~\cite{Cheng:2013xha,Cheng:2013eu,Cheng:2013bca} we expect the FP to be closest to the $\beta=5.5-6.0$ range, so that the dependence on $\beta$ should be weakest in this range.

We perform a combined fit versus  $\beta$ and $t$  using common $\gamma_0$, $\alpha_1$ and $\alpha_2$, but allowing $\beta$ dependent coefficients $c_\beta$ and $d_\beta$. The central fit, as shown in \fig{fig:gamma-M}, omits $\beta=4.0$ and discards the smallest and two largest $t$ values, predicting $\gamma_m=0.23$.  The other exponents obtained are $\alpha_1 = -0.25(14)$ and $\alpha_2 = -2.37(29)$; these likely include some remaining finite-volume effects and thus should not correspond directly to irrelevant operator dimensions.

We vary the analysis by dropping small/large $t$ values, and also including or discarding $\beta=4.0$ and $\beta=6.0$ from the fit; from these variations we estimate a systematic error of $0.04$ on $\gamma_m$.  As an additional cross-check on our finite volume correction procedure, we perform an alternative analysis in which a global fit to $\Rop_\op(t)$ is carried out assuming power-law dependence on the dimensionless ratio $\sqrt{8t}/L$.  This gives a central value of 0.27.  We conservatively take the difference in central values as an estimate of our finite-volume extrapolation systematic, giving the final prediction
\begin{equation}
\gamma_m = 0.23(6)
\end{equation}
combining the systematic errors in quadrature. 

\begin{figure}
\includegraphics[width=0.48\textwidth]{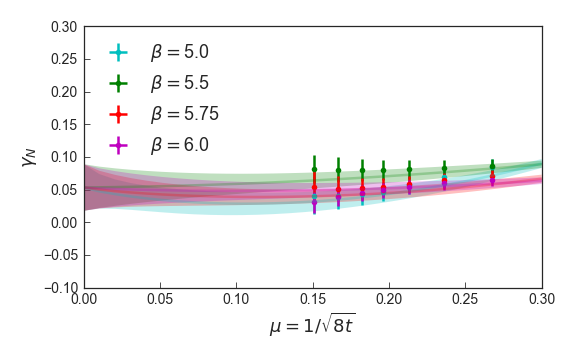}
\caption{Extrapolation of the nucleon anomalous dimension $\gamma_N$ to the infrared limit, as described in the text.  \label{fig:gamma-N}}
\end{figure}

A significant advantage of this technique is that more complicated composite operators can be dealt with in a straightforward way.  To demonstrate this, we consider the nucleon operator with our method.  The nucleon shows more significant oscillations in the ratio $\Rop_N$, continuing into the plateau region; we account for the oscillations by averaging over adjacent pairs of $x_0$ values to obtain $\Rop_N$.  The oscillations at large $x_0$ may be due to the coupling of the staggered nucleon operator to other wrong-parity states; numerically the coupling is small in the ratio.  We define the nucleon anomalous dimension  with an additional negative sign, $\gamma_N \equiv \Delta_N - d_N$, to match the convention of \refcite{Pica:2016rmv,Gracey:2018oym}. Repeating the full analysis as described yields \fig{fig:gamma-N} and predicts
\begin{equation}
\gamma_N = 0.05(5)
\end{equation}
where the finite-volume systematic error is estimated to be 0.03 and the remaining combined systematic and statistical error is 0.04.

\paragraph{Conclusion --} 

We have shown that gradient flow (GF) can be used to study renormalization group (RG) transformations directly, with no need for costly ensemble matching, yielding significantly higher statistical precision and lower cost than conventional MCRG techniques.  The use of correlation functions at distances $x_0 \gg \sqrt{t}$ is crucial for our construction, as is the use of a conserved current to explicitly account for wavefunction renormalization under RG, without which the GF transformation does not have a fixed point.  We have worked effectively at zero fermion mass, but it would be interesting to consider whether extrapolation from $m \neq 0$ in the vicinity of a fixed point would be practical in future work.  There are many possibilities for future studies of other conformal theories using this technique, such as $\mathcal{N}=4$ super-Yang-Mills \cite{Schaich:2015daa} or $\phi^4$ theory in three dimensions.


Finally, it would be very interesting if our derivation could be generalized to QCD-like theories in which there is no IR-stable fixed point to work around.  Such an extension could provide a new and general method for operator renormalization in lattice QCD.

\begin{acknowledgments}
\paragraph{Acknowledgments --} A.H. thanks Luigi del Debbio, Chris Monahan, Andreas Schindler, Georg Bergner and the participants of the workshop ``Numerical approaches to 
holography, quantum gravity and cosmology", 
 Edinburgh,  for valuable comments and useful discussions. We are grateful to Evan Weinberg for his help with the flowed spectrum measurement. Computations for this work were carried out with resources provided by the USQCD Collaboration, which is funded by the Office of Science of the U.~S.~Department of Energy.  This work has been supported by the U.~S.~Department of Energy under grant number DE-SC0010005.  Brookhaven National Laboratory is supported by the U.~S.~Department of Energy under contract DE-SC0012704.
\end{acknowledgments}


\bibliography{../../References/RG_GF.bib}

\begin{thebibliography}{48}%
\makeatletter
\providecommand \@ifxundefined [1]{%
 \@ifx{#1\undefined}
}%
\providecommand \@ifnum [1]{%
 \ifnum #1\expandafter \@firstoftwo
 \else \expandafter \@secondoftwo
 \fi
}%
\providecommand \@ifx [1]{%
 \ifx #1\expandafter \@firstoftwo
 \else \expandafter \@secondoftwo
 \fi
}%
\providecommand \natexlab [1]{#1}%
\providecommand \enquote  [1]{``#1''}%
\providecommand \bibnamefont  [1]{#1}%
\providecommand \bibfnamefont [1]{#1}%
\providecommand \citenamefont [1]{#1}%
\providecommand \href@noop [0]{\@secondoftwo}%
\providecommand \href [0]{\begingroup \@sanitize@url \@href}%
\providecommand \@href[1]{\@@startlink{#1}\@@href}%
\providecommand \@@href[1]{\endgroup#1\@@endlink}%
\providecommand \@sanitize@url [0]{\catcode `\\12\catcode `\$12\catcode
  `\&12\catcode `\#12\catcode `\^12\catcode `\_12\catcode `\%12\relax}%
\providecommand \@@startlink[1]{}%
\providecommand \@@endlink[0]{}%
\providecommand \url  [0]{\begingroup\@sanitize@url \@url }%
\providecommand \@url [1]{\endgroup\@href {#1}{\urlprefix }}%
\providecommand \urlprefix  [0]{URL }%
\providecommand \Eprint [0]{\href }%
\providecommand \doibase [0]{http://dx.doi.org/}%
\providecommand \selectlanguage [0]{\@gobble}%
\providecommand \bibinfo  [0]{\@secondoftwo}%
\providecommand \bibfield  [0]{\@secondoftwo}%
\providecommand \translation [1]{[#1]}%
\providecommand \BibitemOpen [0]{}%
\providecommand \bibitemStop [0]{}%
\providecommand \bibitemNoStop [0]{.\EOS\space}%
\providecommand \EOS [0]{\spacefactor3000\relax}%
\providecommand \BibitemShut  [1]{\csname bibitem#1\endcsname}%
\let\auto@bib@innerbib\@empty
\bibitem [{\citenamefont {El-Showk}\ \emph {et~al.}(2012)\citenamefont
  {El-Showk}, \citenamefont {Paulos}, \citenamefont {Poland}, \citenamefont
  {Rychkov}, \citenamefont {Simmons-Duffin},\ and\ \citenamefont
  {Vichi}}]{ElShowk:2012ht}%
  \BibitemOpen
  \bibfield  {author} {\bibinfo {author} {\bibfnamefont {S.}~\bibnamefont
  {El-Showk}}, \bibinfo {author} {\bibfnamefont {M.~F.}\ \bibnamefont
  {Paulos}}, \bibinfo {author} {\bibfnamefont {D.}~\bibnamefont {Poland}},
  \bibinfo {author} {\bibfnamefont {S.}~\bibnamefont {Rychkov}}, \bibinfo
  {author} {\bibfnamefont {D.}~\bibnamefont {Simmons-Duffin}}, \ and\ \bibinfo
  {author} {\bibfnamefont {A.}~\bibnamefont {Vichi}},\ }\href {\doibase
  10.1103/PhysRevD.86.025022} {\bibfield  {journal} {\bibinfo  {journal} {Phys.
  Rev.}\ }\textbf {\bibinfo {volume} {D86}},\ \bibinfo {pages} {025022}
  (\bibinfo {year} {2012})},\ \Eprint {http://arxiv.org/abs/1203.6064}
  {arXiv:1203.6064 [hep-th]} \BibitemShut {NoStop}%
\bibitem [{\citenamefont {Poland}\ and\ \citenamefont
  {Simmons-Duffin}(2016)}]{Poland:2016chs}%
  \BibitemOpen
  \bibfield  {author} {\bibinfo {author} {\bibfnamefont {D.}~\bibnamefont
  {Poland}}\ and\ \bibinfo {author} {\bibfnamefont {D.}~\bibnamefont
  {Simmons-Duffin}},\ }\href {\doibase 10.1038/nphys3761} {\bibfield  {journal}
  {\bibinfo  {journal} {Nature Phys.}\ }\textbf {\bibinfo {volume} {12}},\
  \bibinfo {pages} {535} (\bibinfo {year} {2016})}\BibitemShut {NoStop}%
\bibitem [{\citenamefont {Poland}\ \emph {et~al.}(2018)\citenamefont {Poland},
  \citenamefont {Rychkov},\ and\ \citenamefont {Vichi}}]{Poland:2018epd}%
  \BibitemOpen
  \bibfield  {author} {\bibinfo {author} {\bibfnamefont {D.}~\bibnamefont
  {Poland}}, \bibinfo {author} {\bibfnamefont {S.}~\bibnamefont {Rychkov}}, \
  and\ \bibinfo {author} {\bibfnamefont {A.}~\bibnamefont {Vichi}},\
  }\href@noop {} {\  (\bibinfo {year} {2018})},\ \Eprint
  {http://arxiv.org/abs/1805.04405} {arXiv:1805.04405 [hep-th]} \BibitemShut
  {NoStop}%
\bibitem [{\citenamefont {Brower}\ \emph {et~al.}(2013)\citenamefont {Brower},
  \citenamefont {Fleming},\ and\ \citenamefont {Neuberger}}]{Brower:2012vg}%
  \BibitemOpen
  \bibfield  {author} {\bibinfo {author} {\bibfnamefont {R.~C.}\ \bibnamefont
  {Brower}}, \bibinfo {author} {\bibfnamefont {G.~T.}\ \bibnamefont {Fleming}},
  \ and\ \bibinfo {author} {\bibfnamefont {H.}~\bibnamefont {Neuberger}},\
  }\href {\doibase 10.1016/j.physletb.2013.03.009} {\bibfield  {journal}
  {\bibinfo  {journal} {Phys. Lett.}\ }\textbf {\bibinfo {volume} {B721}},\
  \bibinfo {pages} {299} (\bibinfo {year} {2013})},\ \Eprint
  {http://arxiv.org/abs/1212.6190} {arXiv:1212.6190 [hep-lat]} \BibitemShut
  {NoStop}%
\bibitem [{\citenamefont {Brower}\ \emph {et~al.}(2017)\citenamefont {Brower},
  \citenamefont {Weinberg}, \citenamefont {Fleming}, \citenamefont {Gasbarro},
  \citenamefont {Raben},\ and\ \citenamefont {Tan}}]{Brower:2016vsl}%
  \BibitemOpen
  \bibfield  {author} {\bibinfo {author} {\bibfnamefont {R.~C.}\ \bibnamefont
  {Brower}}, \bibinfo {author} {\bibfnamefont {E.~S.}\ \bibnamefont
  {Weinberg}}, \bibinfo {author} {\bibfnamefont {G.~T.}\ \bibnamefont
  {Fleming}}, \bibinfo {author} {\bibfnamefont {A.~D.}\ \bibnamefont
  {Gasbarro}}, \bibinfo {author} {\bibfnamefont {T.~G.}\ \bibnamefont {Raben}},
  \ and\ \bibinfo {author} {\bibfnamefont {C.-I.}\ \bibnamefont {Tan}},\ }\href
  {\doibase 10.1103/PhysRevD.95.114510} {\bibfield  {journal} {\bibinfo
  {journal} {Phys. Rev.}\ }\textbf {\bibinfo {volume} {D95}},\ \bibinfo {pages}
  {114510} (\bibinfo {year} {2017})},\ \Eprint
  {http://arxiv.org/abs/1610.08587} {arXiv:1610.08587 [hep-lat]} \BibitemShut
  {NoStop}%
\bibitem [{\citenamefont {Brower}\ \emph {et~al.}(2018)\citenamefont {Brower},
  \citenamefont {Cheng}, \citenamefont {Weinberg}, \citenamefont {Fleming},
  \citenamefont {Gasbarro}, \citenamefont {Raben},\ and\ \citenamefont
  {Tan}}]{Brower:2018szu}%
  \BibitemOpen
  \bibfield  {author} {\bibinfo {author} {\bibfnamefont {R.~C.}\ \bibnamefont
  {Brower}}, \bibinfo {author} {\bibfnamefont {M.}~\bibnamefont {Cheng}},
  \bibinfo {author} {\bibfnamefont {E.~S.}\ \bibnamefont {Weinberg}}, \bibinfo
  {author} {\bibfnamefont {G.~T.}\ \bibnamefont {Fleming}}, \bibinfo {author}
  {\bibfnamefont {A.~D.}\ \bibnamefont {Gasbarro}}, \bibinfo {author}
  {\bibfnamefont {T.~G.}\ \bibnamefont {Raben}}, \ and\ \bibinfo {author}
  {\bibfnamefont {C.-I.}\ \bibnamefont {Tan}},\ }\href {\doibase
  10.1103/PhysRevD.98.014502} {\bibfield  {journal} {\bibinfo  {journal} {Phys.
  Rev.}\ }\textbf {\bibinfo {volume} {D98}},\ \bibinfo {pages} {014502}
  (\bibinfo {year} {2018})},\ \Eprint {http://arxiv.org/abs/1803.08512}
  {arXiv:1803.08512 [hep-lat]} \BibitemShut {NoStop}%
\bibitem [{\citenamefont {Swendsen}(1979)}]{Swendsen:1979gn}%
  \BibitemOpen
  \bibfield  {author} {\bibinfo {author} {\bibfnamefont {R.~H.}\ \bibnamefont
  {Swendsen}},\ }\href {\doibase 10.1103/PhysRevLett.42.859} {\bibfield
  {journal} {\bibinfo  {journal} {Phys. Rev. Lett.}\ }\textbf {\bibinfo
  {volume} {42}},\ \bibinfo {pages} {859} (\bibinfo {year} {1979})}\BibitemShut
  {NoStop}%
\bibitem [{\citenamefont {Bowler}\ \emph {et~al.}(1985)\citenamefont {Bowler},
  \citenamefont {Kenway}, \citenamefont {Pawley}, \citenamefont {Wallace},
  \citenamefont {Hasenfratz}, \citenamefont {Hasenfratz}, \citenamefont
  {Heller}, \citenamefont {Karsch},\ and\ \citenamefont
  {Montvay}}]{Bowler:1984hv}%
  \BibitemOpen
  \bibfield  {author} {\bibinfo {author} {\bibfnamefont {K.~C.}\ \bibnamefont
  {Bowler}}, \bibinfo {author} {\bibfnamefont {R.~D.}\ \bibnamefont {Kenway}},
  \bibinfo {author} {\bibfnamefont {G.~S.}\ \bibnamefont {Pawley}}, \bibinfo
  {author} {\bibfnamefont {D.~J.}\ \bibnamefont {Wallace}}, \bibinfo {author}
  {\bibfnamefont {A.}~\bibnamefont {Hasenfratz}}, \bibinfo {author}
  {\bibfnamefont {P.}~\bibnamefont {Hasenfratz}}, \bibinfo {author}
  {\bibfnamefont {U.~M.}\ \bibnamefont {Heller}}, \bibinfo {author}
  {\bibfnamefont {F.}~\bibnamefont {Karsch}}, \ and\ \bibinfo {author}
  {\bibfnamefont {I.}~\bibnamefont {Montvay}},\ }\href {\doibase
  10.1016/0550-3213(85)90341-4} {\bibfield  {journal} {\bibinfo  {journal}
  {Nucl. Phys.}\ }\textbf {\bibinfo {volume} {B257}},\ \bibinfo {pages} {155}
  (\bibinfo {year} {1985})}\BibitemShut {NoStop}%
\bibitem [{\citenamefont {Hasenfratz}(2012)}]{Hasenfratz:2011xn}%
  \BibitemOpen
  \bibfield  {author} {\bibinfo {author} {\bibfnamefont {A.}~\bibnamefont
  {Hasenfratz}},\ }\href {\doibase 10.1103/PhysRevLett.108.061601} {\bibfield
  {journal} {\bibinfo  {journal} {Phys. Rev. Lett.}\ }\textbf {\bibinfo
  {volume} {108}},\ \bibinfo {pages} {061601} (\bibinfo {year} {2012})},\
  \Eprint {http://arxiv.org/abs/1106.5293} {arXiv:1106.5293 [hep-lat]}
  \BibitemShut {NoStop}%
\bibitem [{\citenamefont {Narayanan}\ and\ \citenamefont
  {Neuberger}(2006)}]{Narayanan:2006rf}%
  \BibitemOpen
  \bibfield  {author} {\bibinfo {author} {\bibfnamefont {R.}~\bibnamefont
  {Narayanan}}\ and\ \bibinfo {author} {\bibfnamefont {H.}~\bibnamefont
  {Neuberger}},\ }\href {\doibase 10.1088/1126-6708/2006/03/064} {\bibfield
  {journal} {\bibinfo  {journal} {JHEP}\ }\textbf {\bibinfo {volume} {03}},\
  \bibinfo {pages} {064} (\bibinfo {year} {2006})},\ \Eprint
  {http://arxiv.org/abs/hep-th/0601210} {arXiv:hep-th/0601210 [hep-th]}
  \BibitemShut {NoStop}%
\bibitem [{\citenamefont {Luscher}(2010{\natexlab{a}})}]{Luscher:2009eq}%
  \BibitemOpen
  \bibfield  {author} {\bibinfo {author} {\bibfnamefont {M.}~\bibnamefont
  {Luscher}},\ }\href {\doibase 10.1007/s00220-009-0953-7} {\bibfield
  {journal} {\bibinfo  {journal} {Commun. Math. Phys.}\ }\textbf {\bibinfo
  {volume} {293}},\ \bibinfo {pages} {899} (\bibinfo {year}
  {2010}{\natexlab{a}})},\ \Eprint {http://arxiv.org/abs/0907.5491}
  {arXiv:0907.5491 [hep-lat]} \BibitemShut {NoStop}%
\bibitem [{\citenamefont {Wilson}\ and\ \citenamefont
  {Kogut}(1974)}]{Wilson:1973jj}%
  \BibitemOpen
  \bibfield  {author} {\bibinfo {author} {\bibfnamefont {K.~G.}\ \bibnamefont
  {Wilson}}\ and\ \bibinfo {author} {\bibfnamefont {J.~B.}\ \bibnamefont
  {Kogut}},\ }\href {\doibase 10.1016/0370-1573(74)90023-4} {\bibfield
  {journal} {\bibinfo  {journal} {Phys. Rept.}\ }\textbf {\bibinfo {volume}
  {12}},\ \bibinfo {pages} {75} (\bibinfo {year} {1974})}\BibitemShut {NoStop}%
\bibitem [{\citenamefont {Polchinski}(1984)}]{Polchinski:1983gv}%
  \BibitemOpen
  \bibfield  {author} {\bibinfo {author} {\bibfnamefont {J.}~\bibnamefont
  {Polchinski}},\ }\href {\doibase 10.1016/0550-3213(84)90287-6} {\bibfield
  {journal} {\bibinfo  {journal} {Nucl. Phys.}\ }\textbf {\bibinfo {volume}
  {B231}},\ \bibinfo {pages} {269} (\bibinfo {year} {1984})}\BibitemShut
  {NoStop}%
\bibitem [{\citenamefont {Berges}\ \emph {et~al.}(2002)\citenamefont {Berges},
  \citenamefont {Tetradis},\ and\ \citenamefont {Wetterich}}]{Berges:2000ew}%
  \BibitemOpen
  \bibfield  {author} {\bibinfo {author} {\bibfnamefont {J.}~\bibnamefont
  {Berges}}, \bibinfo {author} {\bibfnamefont {N.}~\bibnamefont {Tetradis}}, \
  and\ \bibinfo {author} {\bibfnamefont {C.}~\bibnamefont {Wetterich}},\ }\href
  {\doibase 10.1016/S0370-1573(01)00098-9} {\bibfield  {journal} {\bibinfo
  {journal} {Phys. Rept.}\ }\textbf {\bibinfo {volume} {363}},\ \bibinfo
  {pages} {223} (\bibinfo {year} {2002})},\ \Eprint
  {http://arxiv.org/abs/hep-ph/0005122} {arXiv:hep-ph/0005122 [hep-ph]}
  \BibitemShut {NoStop}%
\bibitem [{\citenamefont {Bagnuls}\ and\ \citenamefont
  {Bervillier}(2001)}]{Bagnuls:2000ae}%
  \BibitemOpen
  \bibfield  {author} {\bibinfo {author} {\bibfnamefont {C.}~\bibnamefont
  {Bagnuls}}\ and\ \bibinfo {author} {\bibfnamefont {C.}~\bibnamefont
  {Bervillier}},\ }\href {\doibase 10.1016/S0370-1573(00)00137-X} {\bibfield
  {journal} {\bibinfo  {journal} {Phys. Rept.}\ }\textbf {\bibinfo {volume}
  {348}},\ \bibinfo {pages} {91} (\bibinfo {year} {2001})},\ \Eprint
  {http://arxiv.org/abs/hep-th/0002034} {arXiv:hep-th/0002034 [hep-th]}
  \BibitemShut {NoStop}%
\bibitem [{\citenamefont {Aoki}\ \emph {et~al.}(2015)\citenamefont {Aoki},
  \citenamefont {Kikuchi},\ and\ \citenamefont {Onogi}}]{Aoki:2014dxa}%
  \BibitemOpen
  \bibfield  {author} {\bibinfo {author} {\bibfnamefont {S.}~\bibnamefont
  {Aoki}}, \bibinfo {author} {\bibfnamefont {K.}~\bibnamefont {Kikuchi}}, \
  and\ \bibinfo {author} {\bibfnamefont {T.}~\bibnamefont {Onogi}},\ }\href
  {\doibase 10.1007/JHEP04(2015)156} {\bibfield  {journal} {\bibinfo  {journal}
  {JHEP}\ }\textbf {\bibinfo {volume} {04}},\ \bibinfo {pages} {156} (\bibinfo
  {year} {2015})},\ \Eprint {http://arxiv.org/abs/1412.8249} {arXiv:1412.8249
  [hep-th]} \BibitemShut {NoStop}%
\bibitem [{\citenamefont {Aoki}\ \emph {et~al.}(2016)\citenamefont {Aoki},
  \citenamefont {Kikuchi},\ and\ \citenamefont {Onogi}}]{Aoki:2015snd}%
  \BibitemOpen
  \bibfield  {author} {\bibinfo {author} {\bibfnamefont {S.}~\bibnamefont
  {Aoki}}, \bibinfo {author} {\bibfnamefont {K.}~\bibnamefont {Kikuchi}}, \
  and\ \bibinfo {author} {\bibfnamefont {T.}~\bibnamefont {Onogi}},\
  }\href@noop {} {\bibfield  {journal} {\bibinfo  {journal} {PoS}\ }\textbf
  {\bibinfo {volume} {LATTICE2015}},\ \bibinfo {pages} {305} (\bibinfo {year}
  {2016})},\ \Eprint {http://arxiv.org/abs/1511.06561} {arXiv:1511.06561
  [hep-lat]} \BibitemShut {NoStop}%
\bibitem [{\citenamefont {Monahan}\ and\ \citenamefont
  {Orginos}(2015)}]{Monahan:2015lha}%
  \BibitemOpen
  \bibfield  {author} {\bibinfo {author} {\bibfnamefont {C.}~\bibnamefont
  {Monahan}}\ and\ \bibinfo {author} {\bibfnamefont {K.}~\bibnamefont
  {Orginos}},\ }\href {\doibase 10.1103/PhysRevD.91.074513} {\bibfield
  {journal} {\bibinfo  {journal} {Phys. Rev.}\ }\textbf {\bibinfo {volume}
  {D91}},\ \bibinfo {pages} {074513} (\bibinfo {year} {2015})},\ \Eprint
  {http://arxiv.org/abs/1501.05348} {arXiv:1501.05348 [hep-lat]} \BibitemShut
  {NoStop}%
\bibitem [{\citenamefont {Fujikawa}(2016)}]{Fujikawa:2016qis}%
  \BibitemOpen
  \bibfield  {author} {\bibinfo {author} {\bibfnamefont {K.}~\bibnamefont
  {Fujikawa}},\ }\href {\doibase 10.1007/JHEP03(2016)021} {\bibfield  {journal}
  {\bibinfo  {journal} {JHEP}\ }\textbf {\bibinfo {volume} {03}},\ \bibinfo
  {pages} {021} (\bibinfo {year} {2016})},\ \Eprint
  {http://arxiv.org/abs/1601.01578} {arXiv:1601.01578 [hep-lat]} \BibitemShut
  {NoStop}%
\bibitem [{\citenamefont {Capponi}\ \emph {et~al.}(2016)\citenamefont
  {Capponi}, \citenamefont {Del~Debbio}, \citenamefont {Ehret}, \citenamefont
  {Pellegrini}, \citenamefont {Portelli},\ and\ \citenamefont
  {Rago}}]{Capponi:2016yjz}%
  \BibitemOpen
  \bibfield  {author} {\bibinfo {author} {\bibfnamefont {F.}~\bibnamefont
  {Capponi}}, \bibinfo {author} {\bibfnamefont {L.}~\bibnamefont {Del~Debbio}},
  \bibinfo {author} {\bibfnamefont {S.}~\bibnamefont {Ehret}}, \bibinfo
  {author} {\bibfnamefont {R.}~\bibnamefont {Pellegrini}}, \bibinfo {author}
  {\bibfnamefont {A.}~\bibnamefont {Portelli}}, \ and\ \bibinfo {author}
  {\bibfnamefont {A.}~\bibnamefont {Rago}},\ }\href@noop {} {\bibfield
  {journal} {\bibinfo  {journal} {PoS}\ }\textbf {\bibinfo {volume}
  {LATTICE2016}},\ \bibinfo {pages} {341} (\bibinfo {year} {2016})},\ \Eprint
  {http://arxiv.org/abs/1612.07721} {arXiv:1612.07721 [hep-lat]} \BibitemShut
  {NoStop}%
\bibitem [{\citenamefont {Makino}\ \emph {et~al.}(2018)\citenamefont {Makino},
  \citenamefont {Morikawa},\ and\ \citenamefont {Suzuki}}]{Makino:2018rys}%
  \BibitemOpen
  \bibfield  {author} {\bibinfo {author} {\bibfnamefont {H.}~\bibnamefont
  {Makino}}, \bibinfo {author} {\bibfnamefont {O.}~\bibnamefont {Morikawa}}, \
  and\ \bibinfo {author} {\bibfnamefont {H.}~\bibnamefont {Suzuki}},\
  }\href@noop {} {\  (\bibinfo {year} {2018})},\ \Eprint
  {http://arxiv.org/abs/1802.07897} {arXiv:1802.07897 [hep-th]} \BibitemShut
  {NoStop}%
\bibitem [{\citenamefont {Abe}\ and\ \citenamefont
  {Fukuma}(2018)}]{Abe:2018zdc}%
  \BibitemOpen
  \bibfield  {author} {\bibinfo {author} {\bibfnamefont {Y.}~\bibnamefont
  {Abe}}\ and\ \bibinfo {author} {\bibfnamefont {M.}~\bibnamefont {Fukuma}},\
  }\href@noop {} {\  (\bibinfo {year} {2018})},\ \Eprint
  {http://arxiv.org/abs/1805.12094} {arXiv:1805.12094 [hep-th]} \BibitemShut
  {NoStop}%
\bibitem [{\citenamefont {Luscher}(2013)}]{Luscher:2013cpa}%
  \BibitemOpen
  \bibfield  {author} {\bibinfo {author} {\bibfnamefont {M.}~\bibnamefont
  {Luscher}},\ }\href {\doibase 10.1007/JHEP04(2013)123} {\bibfield  {journal}
  {\bibinfo  {journal} {JHEP}\ }\textbf {\bibinfo {volume} {04}},\ \bibinfo
  {pages} {123} (\bibinfo {year} {2013})},\ \Eprint
  {http://arxiv.org/abs/1302.5246} {arXiv:1302.5246 [hep-lat]} \BibitemShut
  {NoStop}%
\bibitem [{\citenamefont {Gukov}(2016)}]{Gukov:2015qea}%
  \BibitemOpen
  \bibfield  {author} {\bibinfo {author} {\bibfnamefont {S.}~\bibnamefont
  {Gukov}},\ }\href {\doibase 10.1007/JHEP01(2016)020} {\bibfield  {journal}
  {\bibinfo  {journal} {JHEP}\ }\textbf {\bibinfo {volume} {01}},\ \bibinfo
  {pages} {020} (\bibinfo {year} {2016})},\ \Eprint
  {http://arxiv.org/abs/1503.01474} {arXiv:1503.01474 [hep-th]} \BibitemShut
  {NoStop}%
\bibitem [{\citenamefont {Barnes}\ \emph {et~al.}(2004)\citenamefont {Barnes},
  \citenamefont {Intriligator}, \citenamefont {Wecht},\ and\ \citenamefont
  {Wright}}]{Barnes:2004jj}%
  \BibitemOpen
  \bibfield  {author} {\bibinfo {author} {\bibfnamefont {E.}~\bibnamefont
  {Barnes}}, \bibinfo {author} {\bibfnamefont {K.~A.}\ \bibnamefont
  {Intriligator}}, \bibinfo {author} {\bibfnamefont {B.}~\bibnamefont {Wecht}},
  \ and\ \bibinfo {author} {\bibfnamefont {J.}~\bibnamefont {Wright}},\ }\href
  {\doibase 10.1016/j.nuclphysb.2004.09.016} {\bibfield  {journal} {\bibinfo
  {journal} {Nucl. Phys.}\ }\textbf {\bibinfo {volume} {B702}},\ \bibinfo
  {pages} {131} (\bibinfo {year} {2004})},\ \Eprint
  {http://arxiv.org/abs/hep-th/0408156} {arXiv:hep-th/0408156 [hep-th]}
  \BibitemShut {NoStop}%
\bibitem [{\citenamefont {Fodor}\ \emph {et~al.}(2011)\citenamefont {Fodor},
  \citenamefont {Holland}, \citenamefont {Kuti}, \citenamefont {Nogradi},
  \citenamefont {Schroeder}, \citenamefont {Holland}, \citenamefont {Kuti},
  \citenamefont {Nogradi},\ and\ \citenamefont {Schroeder}}]{Fodor:2011tu}%
  \BibitemOpen
  \bibfield  {author} {\bibinfo {author} {\bibfnamefont {Z.}~\bibnamefont
  {Fodor}}, \bibinfo {author} {\bibfnamefont {K.}~\bibnamefont {Holland}},
  \bibinfo {author} {\bibfnamefont {J.}~\bibnamefont {Kuti}}, \bibinfo {author}
  {\bibfnamefont {D.}~\bibnamefont {Nogradi}}, \bibinfo {author} {\bibfnamefont
  {C.}~\bibnamefont {Schroeder}}, \bibinfo {author} {\bibfnamefont
  {K.}~\bibnamefont {Holland}}, \bibinfo {author} {\bibfnamefont
  {J.}~\bibnamefont {Kuti}}, \bibinfo {author} {\bibfnamefont {D.}~\bibnamefont
  {Nogradi}}, \ and\ \bibinfo {author} {\bibfnamefont {C.}~\bibnamefont
  {Schroeder}},\ }\href {\doibase 10.1016/j.physletb.2011.07.037} {\bibfield
  {journal} {\bibinfo  {journal} {Phys. Lett.}\ }\textbf {\bibinfo {volume}
  {B703}},\ \bibinfo {pages} {348} (\bibinfo {year} {2011})},\ \Eprint
  {http://arxiv.org/abs/1104.3124} {arXiv:1104.3124 [hep-lat]} \BibitemShut
  {NoStop}%
\bibitem [{\citenamefont {Cheng}\ \emph
  {et~al.}(2014{\natexlab{a}})\citenamefont {Cheng}, \citenamefont
  {Hasenfratz}, \citenamefont {Liu}, \citenamefont {Petropoulos},\ and\
  \citenamefont {Schaich}}]{Cheng:2013xha}%
  \BibitemOpen
  \bibfield  {author} {\bibinfo {author} {\bibfnamefont {A.}~\bibnamefont
  {Cheng}}, \bibinfo {author} {\bibfnamefont {A.}~\bibnamefont {Hasenfratz}},
  \bibinfo {author} {\bibfnamefont {Y.}~\bibnamefont {Liu}}, \bibinfo {author}
  {\bibfnamefont {G.}~\bibnamefont {Petropoulos}}, \ and\ \bibinfo {author}
  {\bibfnamefont {D.}~\bibnamefont {Schaich}},\ }\href {\doibase
  10.1103/PhysRevD.90.014509} {\bibfield  {journal} {\bibinfo  {journal} {Phys.
  Rev.}\ }\textbf {\bibinfo {volume} {D90}},\ \bibinfo {pages} {014509}
  (\bibinfo {year} {2014}{\natexlab{a}})},\ \Eprint
  {http://arxiv.org/abs/1401.0195} {arXiv:1401.0195 [hep-lat]} \BibitemShut
  {NoStop}%
\bibitem [{\citenamefont {Lombardo}\ \emph {et~al.}(2014)\citenamefont
  {Lombardo}, \citenamefont {Miura}, \citenamefont {Nunes~da Silva},\ and\
  \citenamefont {Pallante}}]{Lombardo:2014pda}%
  \BibitemOpen
  \bibfield  {author} {\bibinfo {author} {\bibfnamefont {M.~P.}\ \bibnamefont
  {Lombardo}}, \bibinfo {author} {\bibfnamefont {K.}~\bibnamefont {Miura}},
  \bibinfo {author} {\bibfnamefont {T.~J.}\ \bibnamefont {Nunes~da Silva}}, \
  and\ \bibinfo {author} {\bibfnamefont {E.}~\bibnamefont {Pallante}},\ }\href
  {\doibase 10.1007/JHEP12(2014)183} {\bibfield  {journal} {\bibinfo  {journal}
  {JHEP}\ }\textbf {\bibinfo {volume} {12}},\ \bibinfo {pages} {183} (\bibinfo
  {year} {2014})},\ \Eprint {http://arxiv.org/abs/1410.0298} {arXiv:1410.0298
  [hep-lat]} \BibitemShut {NoStop}%
\bibitem [{\citenamefont {Lin}\ \emph {et~al.}(2015)\citenamefont {Lin},
  \citenamefont {Ogawa},\ and\ \citenamefont {Ramos}}]{Lin:2015zpa}%
  \BibitemOpen
  \bibfield  {author} {\bibinfo {author} {\bibfnamefont {C.~J.~D.}\
  \bibnamefont {Lin}}, \bibinfo {author} {\bibfnamefont {K.}~\bibnamefont
  {Ogawa}}, \ and\ \bibinfo {author} {\bibfnamefont {A.}~\bibnamefont
  {Ramos}},\ }\href {\doibase 10.1007/JHEP12(2015)103} {\bibfield  {journal}
  {\bibinfo  {journal} {JHEP}\ }\textbf {\bibinfo {volume} {12}},\ \bibinfo
  {pages} {103} (\bibinfo {year} {2015})},\ \Eprint
  {http://arxiv.org/abs/1510.05755} {arXiv:1510.05755 [hep-lat]} \BibitemShut
  {NoStop}%
\bibitem [{\citenamefont {Fodor}\ \emph {et~al.}(2016)\citenamefont {Fodor},
  \citenamefont {Holland}, \citenamefont {Kuti}, \citenamefont {Mondal},
  \citenamefont {Nogradi},\ and\ \citenamefont {Wong}}]{Fodor:2016zil}%
  \BibitemOpen
  \bibfield  {author} {\bibinfo {author} {\bibfnamefont {Z.}~\bibnamefont
  {Fodor}}, \bibinfo {author} {\bibfnamefont {K.}~\bibnamefont {Holland}},
  \bibinfo {author} {\bibfnamefont {J.}~\bibnamefont {Kuti}}, \bibinfo {author}
  {\bibfnamefont {S.}~\bibnamefont {Mondal}}, \bibinfo {author} {\bibfnamefont
  {D.}~\bibnamefont {Nogradi}}, \ and\ \bibinfo {author} {\bibfnamefont
  {C.~H.}\ \bibnamefont {Wong}},\ }\href {\doibase 10.1103/PhysRevD.94.091501}
  {\bibfield  {journal} {\bibinfo  {journal} {Phys. Rev.}\ }\textbf {\bibinfo
  {volume} {D94}},\ \bibinfo {pages} {091501} (\bibinfo {year} {2016})},\
  \Eprint {http://arxiv.org/abs/1607.06121} {arXiv:1607.06121 [hep-lat]}
  \BibitemShut {NoStop}%
\bibitem [{\citenamefont {Hasenfratz}\ and\ \citenamefont
  {Schaich}(2018)}]{Hasenfratz:2016dou}%
  \BibitemOpen
  \bibfield  {author} {\bibinfo {author} {\bibfnamefont {A.}~\bibnamefont
  {Hasenfratz}}\ and\ \bibinfo {author} {\bibfnamefont {D.}~\bibnamefont
  {Schaich}},\ }\href {\doibase 10.1007/JHEP02(2018)132} {\bibfield  {journal}
  {\bibinfo  {journal} {JHEP}\ }\textbf {\bibinfo {volume} {02}},\ \bibinfo
  {pages} {132} (\bibinfo {year} {2018})},\ \Eprint
  {http://arxiv.org/abs/1610.10004} {arXiv:1610.10004 [hep-lat]} \BibitemShut
  {NoStop}%
\bibitem [{\citenamefont {Fodor}\ \emph {et~al.}(2018)\citenamefont {Fodor},
  \citenamefont {Holland}, \citenamefont {Kuti}, \citenamefont {Nogradi},\ and\
  \citenamefont {Wong}}]{Fodor:2017gtj}%
  \BibitemOpen
  \bibfield  {author} {\bibinfo {author} {\bibfnamefont {Z.}~\bibnamefont
  {Fodor}}, \bibinfo {author} {\bibfnamefont {K.}~\bibnamefont {Holland}},
  \bibinfo {author} {\bibfnamefont {J.}~\bibnamefont {Kuti}}, \bibinfo {author}
  {\bibfnamefont {D.}~\bibnamefont {Nogradi}}, \ and\ \bibinfo {author}
  {\bibfnamefont {C.~H.}\ \bibnamefont {Wong}},\ }\href {\doibase
  10.1016/j.physletb.2018.02.008} {\bibfield  {journal} {\bibinfo  {journal}
  {Phys. Lett.}\ }\textbf {\bibinfo {volume} {B779}},\ \bibinfo {pages} {230}
  (\bibinfo {year} {2018})},\ \Eprint {http://arxiv.org/abs/1710.09262}
  {arXiv:1710.09262 [hep-lat]} \BibitemShut {NoStop}%
\bibitem [{\citenamefont {Ryttov}\ and\ \citenamefont
  {Shrock}(2011)}]{Ryttov:2010iz}%
  \BibitemOpen
  \bibfield  {author} {\bibinfo {author} {\bibfnamefont {T.~A.}\ \bibnamefont
  {Ryttov}}\ and\ \bibinfo {author} {\bibfnamefont {R.}~\bibnamefont
  {Shrock}},\ }\href {\doibase 10.1103/PhysRevD.83.056011} {\bibfield
  {journal} {\bibinfo  {journal} {Phys. Rev.}\ }\textbf {\bibinfo {volume}
  {D83}},\ \bibinfo {pages} {056011} (\bibinfo {year} {2011})},\ \Eprint
  {http://arxiv.org/abs/1011.4542} {arXiv:1011.4542 [hep-ph]} \BibitemShut
  {NoStop}%
\bibitem [{\citenamefont {Pica}\ and\ \citenamefont
  {Sannino}(2016)}]{Pica:2016rmv}%
  \BibitemOpen
  \bibfield  {author} {\bibinfo {author} {\bibfnamefont {C.}~\bibnamefont
  {Pica}}\ and\ \bibinfo {author} {\bibfnamefont {F.}~\bibnamefont {Sannino}},\
  }\href {\doibase 10.1103/PhysRevD.94.071702} {\bibfield  {journal} {\bibinfo
  {journal} {Phys. Rev.}\ }\textbf {\bibinfo {volume} {D94}},\ \bibinfo {pages}
  {071702} (\bibinfo {year} {2016})},\ \Eprint
  {http://arxiv.org/abs/1604.02572} {arXiv:1604.02572 [hep-ph]} \BibitemShut
  {NoStop}%
\bibitem [{\citenamefont {Ryttov}\ and\ \citenamefont
  {Shrock}(2017)}]{Ryttov:2017kmx}%
  \BibitemOpen
  \bibfield  {author} {\bibinfo {author} {\bibfnamefont {T.~A.}\ \bibnamefont
  {Ryttov}}\ and\ \bibinfo {author} {\bibfnamefont {R.}~\bibnamefont
  {Shrock}},\ }\href {\doibase 10.1103/PhysRevD.95.105004} {\bibfield
  {journal} {\bibinfo  {journal} {Phys. Rev.}\ }\textbf {\bibinfo {volume}
  {D95}},\ \bibinfo {pages} {105004} (\bibinfo {year} {2017})},\ \Eprint
  {http://arxiv.org/abs/1703.08558} {arXiv:1703.08558 [hep-th]} \BibitemShut
  {NoStop}%
\bibitem [{\citenamefont {Gracey}\ \emph {et~al.}(2018)\citenamefont {Gracey},
  \citenamefont {Ryttov},\ and\ \citenamefont {Shrock}}]{Gracey:2018oym}%
  \BibitemOpen
  \bibfield  {author} {\bibinfo {author} {\bibfnamefont {J.~A.}\ \bibnamefont
  {Gracey}}, \bibinfo {author} {\bibfnamefont {T.~A.}\ \bibnamefont {Ryttov}},
  \ and\ \bibinfo {author} {\bibfnamefont {R.}~\bibnamefont {Shrock}},\
  }\href@noop {} {\  (\bibinfo {year} {2018})},\ \Eprint
  {http://arxiv.org/abs/1805.02729} {arXiv:1805.02729 [hep-th]} \BibitemShut
  {NoStop}%
\bibitem [{\citenamefont {Cheng}\ \emph
  {et~al.}(2014{\natexlab{b}})\citenamefont {Cheng}, \citenamefont
  {Hasenfratz}, \citenamefont {Petropoulos},\ and\ \citenamefont
  {Schaich}}]{Cheng:2013bca}%
  \BibitemOpen
  \bibfield  {author} {\bibinfo {author} {\bibfnamefont {A.}~\bibnamefont
  {Cheng}}, \bibinfo {author} {\bibfnamefont {A.}~\bibnamefont {Hasenfratz}},
  \bibinfo {author} {\bibfnamefont {G.}~\bibnamefont {Petropoulos}}, \ and\
  \bibinfo {author} {\bibfnamefont {D.}~\bibnamefont {Schaich}},\ }\href@noop
  {} {\bibfield  {journal} {\bibinfo  {journal} {PoS}\ }\textbf {\bibinfo
  {volume} {LATTICE2013}},\ \bibinfo {pages} {088} (\bibinfo {year}
  {2014}{\natexlab{b}})},\ \Eprint {http://arxiv.org/abs/1311.1287}
  {arXiv:1311.1287 [hep-lat]} \BibitemShut {NoStop}%
\bibitem [{\citenamefont {Giedt}(2016)}]{Giedt:2015alr}%
  \BibitemOpen
  \bibfield  {author} {\bibinfo {author} {\bibfnamefont {J.}~\bibnamefont
  {Giedt}},\ }\href {\doibase 10.1142/S0217751X16300118} {\bibfield  {journal}
  {\bibinfo  {journal} {Int. J. Mod. Phys.}\ }\textbf {\bibinfo {volume}
  {A31}},\ \bibinfo {pages} {1630011} (\bibinfo {year} {2016})},\ \Eprint
  {http://arxiv.org/abs/1512.09330} {arXiv:1512.09330 [hep-lat]} \BibitemShut
  {NoStop}%
\bibitem [{\citenamefont {Luscher}(2010{\natexlab{b}})}]{Luscher:2010iy}%
  \BibitemOpen
  \bibfield  {author} {\bibinfo {author} {\bibfnamefont {M.}~\bibnamefont
  {Luscher}},\ }\href {\doibase 10.1007/JHEP08(2010)071,
  10.1007/JHEP03(2014)092} {\bibfield  {journal} {\bibinfo  {journal} {JHEP}\
  }\textbf {\bibinfo {volume} {08}},\ \bibinfo {pages} {071} (\bibinfo {year}
  {2010}{\natexlab{b}})},\ \bibinfo {note} {[Erratum: JHEP03,092(2014)]},\
  \Eprint {http://arxiv.org/abs/1006.4518} {arXiv:1006.4518 [hep-lat]}
  \BibitemShut {NoStop}%
\bibitem [{\citenamefont {Cardy}\ \emph {et~al.}(1996)\citenamefont {Cardy},
  \citenamefont {Goddard},\ and\ \citenamefont {Yeomans}}]{cardy1996scaling}%
  \BibitemOpen
  \bibfield  {author} {\bibinfo {author} {\bibfnamefont {J.}~\bibnamefont
  {Cardy}}, \bibinfo {author} {\bibfnamefont {P.}~\bibnamefont {Goddard}}, \
  and\ \bibinfo {author} {\bibfnamefont {J.}~\bibnamefont {Yeomans}},\
  }\href@noop {} {\emph {\bibinfo {title} {{Scaling and Renormalization in
  Statistical Physics}}}},\ Cambridge Lecture Notes in Physics\ (\bibinfo
  {publisher} {Cambridge University Press},\ \bibinfo {year} {1996})\
  p.~\bibinfo {pages} {50}\BibitemShut {NoStop}%
\bibitem [{\citenamefont {Del~Debbio}\ and\ \citenamefont
  {Zwicky}(2010)}]{DelDebbio:2010ze}%
  \BibitemOpen
  \bibfield  {author} {\bibinfo {author} {\bibfnamefont {L.}~\bibnamefont
  {Del~Debbio}}\ and\ \bibinfo {author} {\bibfnamefont {R.}~\bibnamefont
  {Zwicky}},\ }\href {\doibase 10.1103/PhysRevD.82.014502} {\bibfield
  {journal} {\bibinfo  {journal} {Phys. Rev.}\ }\textbf {\bibinfo {volume}
  {D82}},\ \bibinfo {pages} {014502} (\bibinfo {year} {2010})},\ \Eprint
  {http://arxiv.org/abs/1005.2371} {arXiv:1005.2371 [hep-ph]} \BibitemShut
  {NoStop}%
\bibitem [{\citenamefont {Hasenfratz}\ and\ \citenamefont
  {Knechtli}(2001)}]{Hasenfratz:2001hp}%
  \BibitemOpen
  \bibfield  {author} {\bibinfo {author} {\bibfnamefont {A.}~\bibnamefont
  {Hasenfratz}}\ and\ \bibinfo {author} {\bibfnamefont {F.}~\bibnamefont
  {Knechtli}},\ }\href {\doibase 10.1103/PhysRevD.64.034504} {\bibfield
  {journal} {\bibinfo  {journal} {Phys. Rev.}\ }\textbf {\bibinfo {volume}
  {D64}},\ \bibinfo {pages} {034504} (\bibinfo {year} {2001})},\ \Eprint
  {http://arxiv.org/abs/hep-lat/0103029} {arXiv:hep-lat/0103029 [hep-lat]}
  \BibitemShut {NoStop}%
\bibitem [{\citenamefont {Hasenfratz}\ \emph {et~al.}(2007)\citenamefont
  {Hasenfratz}, \citenamefont {Hoffmann},\ and\ \citenamefont
  {Schaefer}}]{Hasenfratz:2007rf}%
  \BibitemOpen
  \bibfield  {author} {\bibinfo {author} {\bibfnamefont {A.}~\bibnamefont
  {Hasenfratz}}, \bibinfo {author} {\bibfnamefont {R.}~\bibnamefont
  {Hoffmann}}, \ and\ \bibinfo {author} {\bibfnamefont {S.}~\bibnamefont
  {Schaefer}},\ }\href {\doibase 10.1088/1126-6708/2007/05/029} {\bibfield
  {journal} {\bibinfo  {journal} {JHEP}\ }\textbf {\bibinfo {volume} {05}},\
  \bibinfo {pages} {029} (\bibinfo {year} {2007})},\ \Eprint
  {http://arxiv.org/abs/hep-lat/0702028} {arXiv:hep-lat/0702028 [hep-lat]}
  \BibitemShut {NoStop}%
\bibitem [{\citenamefont {Cheng}\ \emph {et~al.}(2012)\citenamefont {Cheng},
  \citenamefont {Hasenfratz},\ and\ \citenamefont {Schaich}}]{Cheng:2011ic}%
  \BibitemOpen
  \bibfield  {author} {\bibinfo {author} {\bibfnamefont {A.}~\bibnamefont
  {Cheng}}, \bibinfo {author} {\bibfnamefont {A.}~\bibnamefont {Hasenfratz}}, \
  and\ \bibinfo {author} {\bibfnamefont {D.}~\bibnamefont {Schaich}},\ }\href
  {\doibase 10.1103/PhysRevD.85.094509} {\bibfield  {journal} {\bibinfo
  {journal} {Phys. Rev.}\ }\textbf {\bibinfo {volume} {D85}},\ \bibinfo {pages}
  {094509} (\bibinfo {year} {2012})},\ \Eprint {http://arxiv.org/abs/1111.2317}
  {arXiv:1111.2317 [hep-lat]} \BibitemShut {NoStop}%
\bibitem [{\citenamefont {Cheng}\ \emph {et~al.}(2013)\citenamefont {Cheng},
  \citenamefont {Hasenfratz}, \citenamefont {Petropoulos},\ and\ \citenamefont
  {Schaich}}]{Cheng:2013eu}%
  \BibitemOpen
  \bibfield  {author} {\bibinfo {author} {\bibfnamefont {A.}~\bibnamefont
  {Cheng}}, \bibinfo {author} {\bibfnamefont {A.}~\bibnamefont {Hasenfratz}},
  \bibinfo {author} {\bibfnamefont {G.}~\bibnamefont {Petropoulos}}, \ and\
  \bibinfo {author} {\bibfnamefont {D.}~\bibnamefont {Schaich}},\ }\href
  {\doibase 10.1007/JHEP07(2013)061} {\bibfield  {journal} {\bibinfo  {journal}
  {JHEP}\ }\textbf {\bibinfo {volume} {07}},\ \bibinfo {pages} {061} (\bibinfo
  {year} {2013})},\ \Eprint {http://arxiv.org/abs/1301.1355} {arXiv:1301.1355
  [hep-lat]} \BibitemShut {NoStop}%
\bibitem [{\citenamefont {Aoki}\ \emph {et~al.}(2000)\citenamefont {Aoki} \emph
  {et~al.}}]{Aoki:1999av}%
  \BibitemOpen
  \bibfield  {author} {\bibinfo {author} {\bibfnamefont {S.}~\bibnamefont
  {Aoki}} \emph {et~al.} (\bibinfo {collaboration} {JLQCD}),\ }\href {\doibase
  10.1103/PhysRevD.62.094501} {\bibfield  {journal} {\bibinfo  {journal} {Phys.
  Rev.}\ }\textbf {\bibinfo {volume} {D62}},\ \bibinfo {pages} {094501}
  (\bibinfo {year} {2000})},\ \Eprint {http://arxiv.org/abs/hep-lat/9912007}
  {arXiv:hep-lat/9912007 [hep-lat]} \BibitemShut {NoStop}%
\bibitem [{\citenamefont {Ishikawa}\ \emph {et~al.}(2014)\citenamefont
  {Ishikawa}, \citenamefont {Iwasaki}, \citenamefont {Nakayama},\ and\
  \citenamefont {Yoshie}}]{Ishikawa:2013tua}%
  \BibitemOpen
  \bibfield  {author} {\bibinfo {author} {\bibfnamefont {K.~I.}\ \bibnamefont
  {Ishikawa}}, \bibinfo {author} {\bibfnamefont {Y.}~\bibnamefont {Iwasaki}},
  \bibinfo {author} {\bibfnamefont {Y.}~\bibnamefont {Nakayama}}, \ and\
  \bibinfo {author} {\bibfnamefont {T.}~\bibnamefont {Yoshie}},\ }\href
  {\doibase 10.1103/PhysRevD.89.114503} {\bibfield  {journal} {\bibinfo
  {journal} {Phys. Rev.}\ }\textbf {\bibinfo {volume} {D89}},\ \bibinfo {pages}
  {114503} (\bibinfo {year} {2014})},\ \Eprint {http://arxiv.org/abs/1310.5049}
  {arXiv:1310.5049 [hep-lat]} \BibitemShut {NoStop}%
\bibitem [{\citenamefont {Schaich}\ and\ \citenamefont
  {Catterall}(2017)}]{Schaich:2015daa}%
  \BibitemOpen
  \bibfield  {author} {\bibinfo {author} {\bibfnamefont {D.}~\bibnamefont
  {Schaich}}\ and\ \bibinfo {author} {\bibfnamefont {S.}~\bibnamefont
  {Catterall}},\ }\href {\doibase 10.1142/S0217751X17470194,
  10.1142/9789813231467_0028, 10.1142/9789813231467_0028
  10.1142/S0217751X17470194} {\bibfield  {journal} {\bibinfo  {journal} {Int.
  J. Mod. Phys.}\ }\textbf {\bibinfo {volume} {A32}},\ \bibinfo {pages}
  {1747019} (\bibinfo {year} {2017})},\ \bibinfo {note} {[,199(2018)]},\
  \Eprint {http://arxiv.org/abs/1508.00884} {arXiv:1508.00884 [hep-th]}
  \BibitemShut {NoStop}%
\end{thebibliography}%

\end{document}